\documentclass[preprint,floatfix,superscriptaddress]{revtex4}
\usepackage{graphicx}
\usepackage{amsmath}
\usepackage{latexsym}
\usepackage{epstopdf}
\usepackage{amsmath}
\usepackage{amssymb,amsmath}
\usepackage[toc,page]{appendix}
\usepackage{color}
\usepackage{float}

\newcommand{\beq}{\begin{equation}}
\newcommand{\eeq}{\end{equation}}

\begin{document}

\title{Long-range temporal correlation in Auditory Brainstem Responses to Spoken Syllable /da/}

\author{Marjan Mozaffarilegha}
\affiliation{Ibn-Sina Multidisciplinary laboratory, Department of Physics, Shahid Beheshti University, Tehran, P.O.Box: 1983969411, Iran}

\author{ S. M. S. Movahed }
\email{m.s.movahed@ipm.ir}
\affiliation{Department of Physics, Shahid Beheshti University, Tehran, P.O.Box: 1983969411, Iran}
\affiliation{Ibn-Sina Multidisciplinary laboratory, Department of Physics, Shahid Beheshti University, Tehran, P.O.Box: 1983969411, Iran}

\vskip 1cm

\begin{abstract}
The speech auditory brainstem response (sABR) is an objective clinical tool to diagnose particular impairments along the auditory brainstem pathways. We explore the scaling behavior of the brainstem in response to synthetic /da/ stimuli using a proposed pipeline including Multifractal Detrended Moving Average Analysis (MFDMA) modified by Singular Value Decomposition. The scaling exponent confirms that all normal sABR are classified into the non-stationary process. The average Hurst exponent is $H=0.77\pm0.12$ at 68\% confidence interval indicating long-range correlation which shows the first universality behavior of sABR. Our findings exhibit that fluctuations in the sABR series are dictated by a mechanism associated with long-term memory of the dynamic of the auditory system in the brainstem level. The $q-$dependency of $h(q)$ demonstrates that underlying data sets have multifractal nature revealing the second universality behavior of the normal sABR samples. Comparing Hurst exponent of original sABR with the results of the corresponding shuffled and surrogate series, we conclude that its multifractality is almost due to the long-range temporal correlations which are devoted to the third universality. Finally, the presence of long-range correlation which is related to the slow timescales in the subcortical level and integration of information in the brainstem network is confirmed.
%PACS:{}
%64.60.al	Fractal and multifractal systems (see also 61.43.Hv Fractals; macroscopic aggregates)
%61.43.Hv	Fractals; macroscopic aggregates (including diffusion-limited aggregates)
%Fractals structure of disordered solids, 61.43.Hv
% structure of disordered solids, 61.43.Hv
%structure and roughness of, 68.35.Ct
% Mechanical properties
%of solid surfaces and interfaces, 68.35.Gy
  
  %Surface patterning, 81.65.Cf
  %Fractals
  %structure of disordered solids, 61.43.Hv
  %Nanoscale pattern formation, 81.16.Rf
  %62.23.St	Complex nanostructures, including patterned or assembled structures
  %Anisotropic media, 81.05.Xj
 % disordered structure, 61.43.-j
  

\end{abstract}
\maketitle

\section{Introduction}\label{intro}

The scalp-recorded Auditory Brainstem Response (ABR)is the most common auditory evoked potential that reflects the dynamics of the large populations of neurons along the auditory brainstem to simple acoustic sounds (e.g., tones, click) \cite{plack2018sense, burkard2007auditory, sampath2016brainstem}. This evoked potential can be used for oto-neurological diagnosis, particularly possible lesions along the auditory brainstem pathways \cite{hall2014introduction, ferraro2018auditory,skoe2017reading}.% The complexities and the effects of inter-subject variations on the encoding of sounds are features of the brainstem processing.

However, the ABR to click sounds cannot anticipate encoding of complex sounds because of the nonlinear dynamic behavior of the auditory system. Therefore, in the recent studies, more complex stimuli such as speech or music have been used to evaluate the behavior of the brainstem to the more complex stimuli \cite{kouni2014novel,skoe2010auditory}. The speech ABR (sABR) comprises transient and sustained responses. Transient and non-periodic characteristics of the stimulus are seen in the transient responses, while sustained time-locked responses including periodic characteristics are shown in the sustained responses \cite{ skoe2010auditory, reichenbach2016auditory, sanfins2017speech}. The sABR signal is a non-invasive and objective tool that is suited for evaluating individuals with developmental and learning problems and also for studying the role of the brainstem in encoding complex sounds \cite{reichenbach2016auditory, sanfins2017speech,tarasenko2017sa62,bellier2015speech,koravand2017speech}.

The sABR signal is mostly contaminated by several types of external artifacts such as muscular movement, electroencephalogram, non-biological noises, and trends. A more general definition of trend is a part of a series representing a pattern or dominant behavior. As an illustration, monotonous and periodic features can be considered as well-known trends \cite{wu2007trend}. Over the past two decades, various methods in time and frequency domains have been used to analyze sABR using peak latency and amplitude, cross-correlation, Fourier analysis, and cross-phasogram \cite{fujihira2015correlations,skoe2011cross}. However, these methods conceal any grasp of the exact dynamical features of the sABR signals. Furthermore, most quantitative methods to evaluate the temporal dynamics of sABR need the sABR series to be stationary in which the mean and variance of the signal do not change with time \cite{parish2004long}. While, the recent studies have described non-linear and non-stationary dynamics of sABR signals, limiting the utility of these methods for examining generic features of sABR signals \cite{mozaffarilegha2016identification,mozaffarilegha2018complexity}. In addition, due to the complex nature of sABR signals and the presence of several types of noises, underlying signals are mimicked by noises and trends. Examining such data based on linear analysis is not reliable encouraging taking into account non-linear methods which are effective ways of explaining these complex relationships. Therefore, it is crucial to implement robust methods for analyzing sABR signals in order to remove the destructive effects of various trends and noises.

The integration of information over large time scales is an essential ingredient of neural networks. The dynamics of this network are usually characterized by slow power-law decay or long-range temporal correlations. The long-range temporal correlations intuitively guarantee some memory about the past neural activities across different cortical and sub-cortical areas. Mathematical description of long-range correlation corresponds to the divergence of correlation function integration when the size of data becomes infinitely long \cite{kantelhardt2009fractal}. It has been exhibited that various time series extracted from biological systems, including electrocardiography (ECG) \cite{kudinov2015mathematical, zhang2017design}, electroencephalography (EEG)\cite{sikdar2018epilepsy, chatterjee2017multifractal, pavlov2018multifractal, francca2018fractal, sikdar2017multifractal, khoshnoud2018functional, chakraborty2016detecting, chiang2016detrended, maity2015multifractal, karperien2016morphology, namazi2016analysis, chakraborty2016comparative, favela2016fractal}, signal neuron discharge \cite{andres2017motion} and human gait \cite{cavanaugh2017multifractality} behave as scale-invariant processes. The power-law scaling studies have shown the \say{universality} of the behaviors in the complex biological systems \cite{balasubramaniam2018factorization, breakspear2017dynamic, sangiuliano2018long, blythe2017long, meisel2017interplay}. The\say{universality} indicates that there are properties for a large portion of systems which are independent of the dynamics of systems. In addition, it represents the fact that, a few essential factors are necessary to determine the scaling exponents of a complex system. Such scaling exponents characterize the behavior of a typical system. Consequently, various systems which may seem to be independent of each other can be classified in a category and they behave in a considerably similar manner \cite{kadanoff1990scaling}. For a typical system experiencing a phase transition, the special value of the parameter at which the system changes its phase is the system's critical point. For systems that exhibit universality, the closer the parameter is to its critical value, the less sensitively the order parameter depends on the details of the system. However, several other studies have emphasized extensive fluctuations in scaling exponents across subjects \cite{borges2017scale}. Similarly, it is revealed that long-range temporal correlations can change dramatically with small changes in the connectivity of the original networks \cite {poil2012critical}. %Obviously, it cannot be mentioned that all power laws are "universal".
So, it motivates us to evaluate whether there is a universality of the behaviors in sABR signals.

The sABR series is contaminated by non-stationary sources including trends and non-biological noises. To achieve reliable results, these spurious effects should be removed from the intrinsic fluctuations. Since there is no universal definition for the trends, various methods have been proposed to eliminate the trends from the underlying time series \cite{wu2007trend}. In order to examine long-range temporal correlation in non-stationary time series, some mathematical and computational methods have been proposed recently. Detrended fluctuation analysis (DFA) is the most well-known nonlinear methods for studying non-linear, non-stationary time series\cite{piskorski2018properties,chen2018non,li2017monitoring}. DFA is a scaling analysis that offers a quantitative parameter to characterize the correlation properties in non-stationary data by detrending the data on various time scales to remove spurious discovery of temporal correlations arising as an artifact of non-stationarity. Therefore, this approach allows us to determine an exact scaling exponent of the time series.

The generalized form of DFA which is known as Multifractal Detrended Fluctuations Analysis (MFDFA) is one of the best-known methods to capture multifractality in series \cite{sharifi2017complexity,chatterjee2017multifractal}, and used in various fields, ranging from cosmic microwave background radiations \cite{movahed2011long}, sunspot fluctuations \cite{movahed2006multifractal,madanchi2017strong}, plasma fluctuations \cite{shaw2017investigation}, astronomy \cite{zunino2014experimental}, economic time series \cite{cao2018multifractal,tiwari2017multifractal,ferreira2017assessment}, music \cite{nag2017music, ghosh2018gestalt}, traffic jamming \cite{qiu2018scaling} to image processing \cite{chen2017fractal, ibrahim2018feature} and biological time series \cite{maity2015multifractal,kudinov2015mathematical,chakraborty2016detecting,costa2018fractal}. However this approach is not suitable to completely remove sinusoidal and power-law trends \cite{hu2001effect,wei2017multifractal,li2018effect}. In the presence of a sinusoidal trend superimposed on the data, the fluctuation functions derived by MFDFA or MFDMA contain at least one cross-over. Therefore, one can not assign a unique scaling exponent to clarify the fractality nature of underlying process \cite{hu2001effect,nagarajan2005minimizing2,nagarajan2005minimizing1,eghdami2017multifractal}.  On the other hand, MFDFA contains discontinuity in its internal algorithm for capturing local trends leading to discrepancy in computed scaling exponents. Therefore, MFDMA has been proposed to quantify the statistical properties of mono (multi) fractal time series  \cite{arianos2007detrending,gu2010detrending,shao2015effects}. In addition, several robust methods have been also proposed to remove trends such as Fourier Detrended Fluctuations Analysis (FDFA) \cite{kiyono2015establishing}, Adaptive Detrending method (AD) \cite{jung2018adaptive}, Singular Value Decomposition (SVD) \cite{nagarajan2005minimizing2,maiorino2017data}, and Empirical Mode Decomposition (EMD)\cite{liu2016fault}. In this study, we have used MFDMA and SVD-MFDMA approaches to remove crossover in our results. The crossover is a changing point in a typical scaling function. More precisely, in the presence of a crossover, one can assign different scaling exponents for two regimes. In our paper, we denote the point where a changing is recognized in the fluctuation function behavior by $s_{\times}$ \cite{kantelhardt2002multifractal}. The SVD method can remove trends corresponding to the sinusoidal trends in the results \cite{hajian2010multifractal,maiorino2017data}.

To the best of our knowledge, the multifractal analysis has not been used on the sABR series in the literature. In this paper, for the first time, MFDMA \cite{shao2015effects} method is used to capture the intrinsic multiscaling dynamics and assessment of universality behavior of sABR signals. We will also utilize SVD detrending algorithm to remove or at least decrease the influence of trends and noises as much as possible \cite{maiorino2017data,pavlov2018detection}. Then, the multifractal behavior of sABR signals is checked and in the case of multifractality, we determine the sources of multifractality based on the comparison the MFDMA results to those obtained via the MFDMA for shuffled and surrogate series. Furthermore, we evaluate the pattern of sABR signals and measure long-range temporal correlations in sABR series which has not been obtained as yet using other methods.

The rest of this paper is organized as follows. In Sec. \ref{method}, the theoretical overview of MFDMA and SVD will be explained in details. In Sec. \ref{res}, experimental results of the multifractal methods on sABR series are discussed. The multifractality of sABR series is assessed in this section. Finally, Sec. \ref{dis}  is devoted to the discussion.

\section{Methods}\label{method}

\subsection{Participants and procedures}

\begin{figure}[H]
\centering
\includegraphics[width=0.8\linewidth]{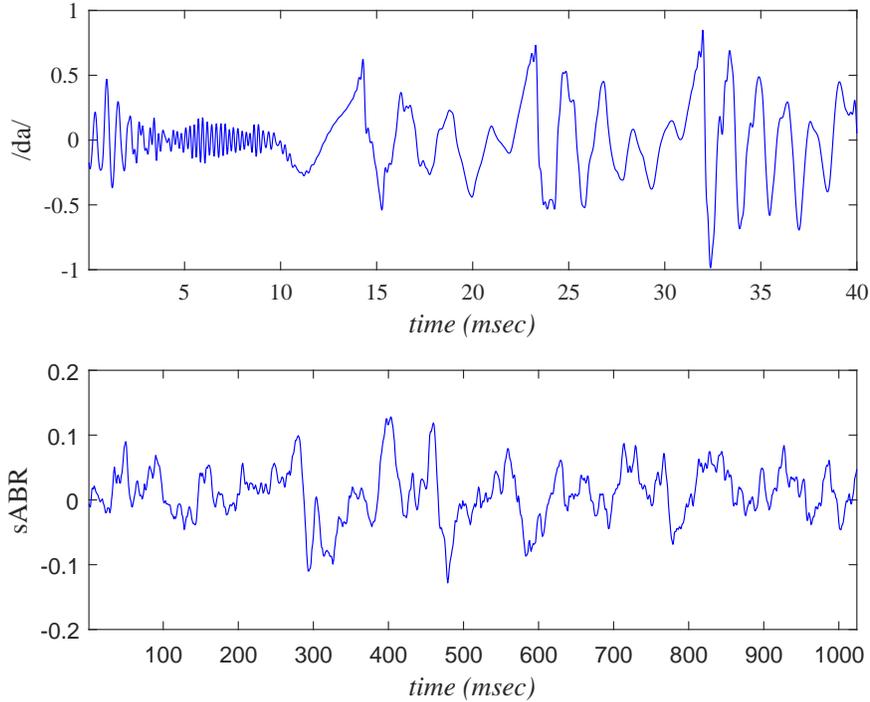}
\caption{A typical /da/ stimulus (upper panel) and corresponding response (sABR) series (lower panel).}
\label{fig:da}
\end{figure}

Forty volunteers from Iran University of Medical Sciences (18 women and 22 men), aged 20-28 years (mean$\pm$SD=22.77$\pm$2.05) contributed to this experiment. Based on their self-report, they were right-handed and monolingual Persian speakers with no history of auditory, learning or neurologic problems. The study has been explained to subjects and then the informed consent has been collected from them. %Consent for participation in this study was signed by subjects.
All procedures have been approved by the deputy of research review board, Iran University of medical sciences. It is worth mentioning that we followed the relevant approved guidelines in this research. The stimuli consisted of a 40 $ms$ speech syllable /da/ provided with Biologic Navigator instrument (Natus Medical Inc., San Carlos, CA, USA). The speech signals are presented to the subjects at a sampling frequency of 44.1 kHz through the computer’s internal sound card. The fundamental frequency ($F_0$) of this syllable is 128 Hz. Complex nature combining both transient and sustained features, the universality of the syllable in most phonetic inventories and its research potential in hearing and learning disorders motivate researchers to utilize Consonant-vowel stimulus /da/ in auditory brainstem study \cite{binkhamis2017methodological}. The contact electrodes were positioned at the vertex ($Cz$) as noninverting, earlobes (inverting) and forehead ($Fpz$) as ground. The stimulus was presented monoaurally at 80 dB nHL via Biologic insert earphone (580-SINSER), with a repetition rate of 7.1/s. The sABR responses were recorded using 1024 digital sampling points over an 85.33 $ms$ time window in the right ear. Finally, the average of total 6000 artifact free responses was collected from every volunteer. Fig. \ref{fig:da} indicates a typical synthesis stimulus of /da/ and sABR signals used for further analysis. It is worth noting that, here we collect s-ABR series in the brainstem level (superior olivary complex and lateral lemniscus) to investigate the brainstem temporal encoding of speech.

\subsection{MFDMA}

The multifractal detrended moving average analysis is a modified version of multifractal detrended fluctuation analysis (MFDFA).
MFDMA method has been introduced to solve the presence of a discontinuity for fitting a polynomial at the boundary of each partition in MFDFA \cite{alessio2002second, shao2015effects}. This algorithm is described as follows:
\label{sec:2}

(1): Consider $sABR(i)$ as a time series where $i=1,\ldots,N$, then, construct the sequence of cumulative sums as follow:
\begin{eqnarray}
X(j)=\sum_{i=1}^{j} [{\it sABR}(i)-\langle {\it sABR}\rangle], \qquad j=1,...,N.
\end{eqnarray}
here $\langle.\rangle$ corresponds to average on available data.

(2): Calculate the moving average function $\widetilde{X(j)}$:
\begin{eqnarray}
\widetilde{X(j)}=\frac{1}{s}\sum_{-\lfloor(s-1)\theta\rfloor}^{\lceil(s-1)(1-\theta)\rceil}X(j-k),
\end{eqnarray}
Where $s$ is the window size, and $\theta$ is the position parameter varying in the range [0,1]. Hence, the moving average $\theta=0$ is called the backward moving average, $\theta=0.5$ corresponds to the centered moving average, and $\theta=1$ refers to the forward moving average \cite{gu2010detrending,shao2015effects}. Throughout this paper we set $\theta=0.0$ due to its robustness reported in various papers \cite{shao2015effects,xu2017direct}

(3): Construct the detrended data by subtracting computed moving average function from the cumulative series $X$ as:
\begin{equation}
\varepsilon(i) \equiv X(i)-\widetilde{X(i)},
\end{equation}
where $s-s_1\leq i\leq N-s_1$

(4): Now, divide $\varepsilon(i)$ into $N_s= int[N/s]$ non-overlapping windows with the size of $s$ and then we have fluctuation function as follows:
\begin{equation}
\mathcal{F}^2(s,\nu) = \frac{1}{s}\sum_{i=1}^{s}\varepsilon^2(i+({\nu}-1)s).
\end{equation}

(5): Calculate the $qth$ order of overall fluctuation function by:
\begin{equation}\label{eq:fluctuation function}
\mathcal{F}_{q}(s) = \left(\frac{1}{N_s}\sum_{\nu=1}^{N_s} [\mathcal{F}^2(s,\nu)]^{q/2}\right)^{\frac{1}{q}},
\end{equation}
for $q =0$ according to L'H\"{o}spital's rule, we have:
\begin{equation}\label{eq:f0}
\mathcal{F}_{0}(s) = \exp\left( \frac{1}{2N_s}\sum_{\nu=1}^{N_s}\ln\mathcal{F}^2(s,\nu)  \right).
\end{equation}
Finally, the scaling form $\mathcal{F}_{q}(s)$ is supposed to be as follows:
\begin{equation}
\mathcal{F}_{q}(s) \sim s^{h(q)}.
\label{eq:h}
\end{equation}
The $h(q)$ is called the generalized Hurst exponent\cite{hurst1951long, xu2017direct,eghdami2017multifractal}. Any $q$-dependency of $h(q)$, represents multifractality in underlying time series. For non-stationary time series, Hurst exponent derived by MFDMA is $h(q=2)>1$, thus in this case Hurst exponent is given by: $H=h(q=2)-1$. For stationary random time series $H=0.5$, while for persistent time series $0.5<H<1.0$. For anti-correlated time series, $H<0.5$ \cite{peng1994mosaic,Taqqu1995,Delignieres2005,xu2017direct,eghdami2017multifractal}. The correlation properties and self-similarity of time series can be measured using Hurst exponent. It can also determine the presence of long range correlation.

To compute the reliable generalized Hurst exponent (equation \ref{eq:h}), generally the likelihood statistics has been implemented \cite{indrayan2017medical,royall2017statistical}. For a Gaussian distribution, $68\%$ confidence interval corresponds to the integration over variable on the interval represented by $[-\sigma, +\sigma]$. In this paper we computed the standard deviation of scaling exponents and reported by $\pm \sigma$ in order to clarify the statistical uncertainty in derived exponents (to make more sense see also \cite{eghdami2017multifractal}).
There is a non-linear dependency between $\tau(q)$ and $q$ in multifractal signals. The Hurst exponent $h(q)$ is related to the scaling exponent $\tau(q)$ by the following formula:
\begin{equation}
\label{eq: tau and h}
\tau(q)=qh(q)-1.
\end{equation}

To characterize multifractality more quantitatively, the so-called singularity spectrum is defined by Legendre transformation as \cite{feder1988fractals, peitgen2006chaos}:
\begin{equation}
f(\alpha)=q[\alpha-h(q)]+1,
\label{eq:falpha}
\end{equation}
and
\begin{equation}
\alpha=h(q)+qh'(q)
\end{equation}
where $\alpha$ is the $h\ddot{o}lder$ exponent and $f(\alpha)$ denotes the dimension of the subset series that is characterized by $\alpha$. In other word, the singularity spectrum reveals a value indicating the scaling behavior of the signal. The domain of H\"{o}lder spectrum, $\alpha\in[\alpha_{min},\alpha_{max}]$, becomes \cite{muzy1994multifractal,xu2017direct,eghdami2017multifractal}:
\begin{equation}
\alpha_{min}=\lim_{q \to +\infty} \frac{\partial \tau (q)}{\partial q},
\end{equation}
\begin{equation}
\alpha_{max}=\lim_{q \to -\infty} \frac{\partial \tau (q)}{\partial q},
\end{equation}
The width of the spectrum, $\Delta\alpha$, gives a measure of the multifractality (complexity) of the spectrum. It can be explained as follows:
\begin{equation}
\label{eq: delta alpha}
\Delta\alpha\equiv\alpha_{max}-\alpha_{min}.
\end{equation}
The higher values of the multifractality nature represent the complexity of underlying time series. The origin of multifractality in a sABR series can be confirmed by evaluating the corresponding shuffled and the surrogate time series. As explained in more details by J.W. Kantelhardt et al.\cite{kantelhardt2002multifractal}, generally, there is two different types of multifractality in a time series: (i) Multifractality due to a broadness probability density function of the time series. In this case, the multifractality of the time series cannot be removed by random shuffling. (ii) Multifractality due to variations of long-range correlations in small and large fluctuations. Here, the probability density function of the time series can have a distribution with finite moments. The shuffling procedure destroys all long-range correlations. In another word, the auto-correlation of a shuffled series behaves as $\langle x(t)x(t')\rangle\sim \delta_{\rm dirak}(t-t')$. Hence, if the multifractality of the original time series belongs to the long-range correlation, the shuffled data will show the non-fractal scaling. However, the multifractality due to the broadness of the probability density function is not affected by the shuffling procedure. To determine the multifractality due to the broadness of the probability density function, the surrogate method is used. If the multifractality of the original time series belongs to a broad probability density function, $h(q)$ obtained by the surrogate method will be independent of $q$. If sABR series has both kinds of multifractality, then the shuffled and surrogate series will have weaker multifractality than the original time series.
\subsection{Singular Value Decomposition (SVD)}
SVD method can be determined in the following steps \cite{nagarajan2005minimizing1, nagarajan2005minimizing2, hajian2010multifractal}:

(1): Construct a matrix which its elements are sABR series with the following order:
\begin{equation}
\gamma(k)=(sABR_k,sABR_{k+\tau},\cdots,sABR_{k+N-(d-1)\tau}),\qquad 1\leq k \leq d.
\end{equation}
\begin{equation}
\Gamma=
 \begin{bmatrix}
    \gamma_1  \\
    .\\
      .\\
    \gamma_d  \\
  \end{bmatrix}.
\end{equation}
where $d$ is embedding dimension, $\tau$ is the time delay. For a time series with length $N$, the maximum value of embedding dimension $d$ is equal to $d\leq N-(d-1)\tau+1$ \cite{nagarajan2005minimizing1, hajian2010multifractal,maiorino2017data}.

(2):Decompose matrix ${{\Gamma}}$ to left (${\mathbf{U}}_{d\times d}$) and right (${\mathbf{V}}_{(N-(d-1)\tau)\times(N-(d-1)\tau)}$) orthogonal matrices:
\begin{equation}
\Gamma=U{\Sigma} V^{\dagger},
\end{equation}
where ${{\Sigma}}_{d\times (N-(d-1)\tau)}$ is a diagonal matrix and its elements are the singular values. We set $2p+1$ the dominant eigenvalues in the matrix ${\Sigma}$ to zero to eliminate long periods. Then, set the dominant eigenvalues in the matrix ${{\Sigma}}$ to zero. The resulting matrix will be ${{\Sigma^*}}$.

(3): Determine the filtered matrix as $\widetilde{\Gamma}=U{\Sigma}^* V^{\dagger}$.

(4): Map back the filtered matrix on to a filtered time series as follows:
\begin{equation}
\widetilde{{\it sABR}}_{i+j-1}=\widetilde{\Gamma}_{ij}.
\end{equation}
where $1\le j\le N-(d-1)\tau$ and $1\le i\le d$. Finally, the cleaned sABR series will be used as input for MFDMA.

\section{Results}\label{res}

In this section, we utilize MFDMA accompanying SVD method to evaluate the multifractal nature and the complexity of sABR data leading to clarify the statistical behavior of observed sABR series.

\begin{figure}[H]
\centering
\includegraphics[width=\linewidth]{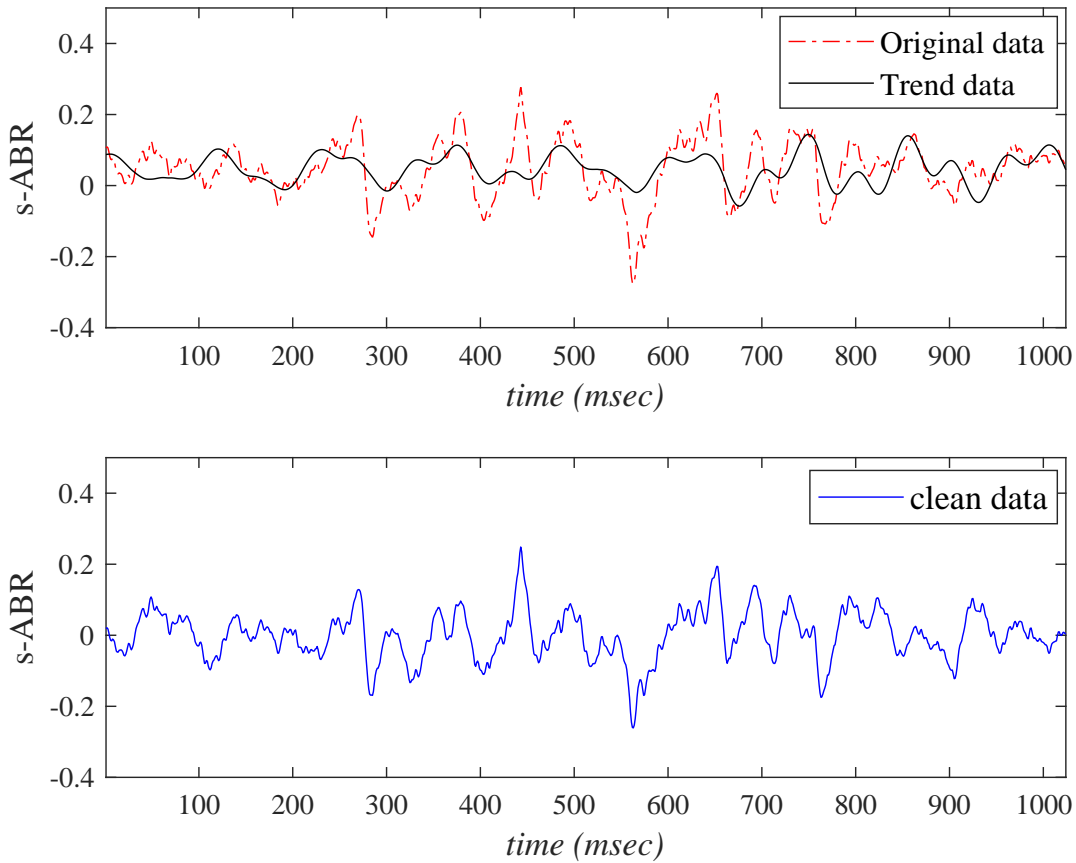}
\label{fig:U}
\caption{{Implementation of SVD on a typical normal sABR signal. The upper panel corresponds to observed data (dash-dot line) and trend (solid line), while the bottom represents the residual data corresponding to clean data.}}
\label{fig:I}
\end{figure}

The upper panel of Fig. \ref{fig:I} depicts a typical normal observed sABR signal and associated trend determined by the SVD method. The corresponding residue between the observed time series and trend is plotted in the lower panel of Fig. \ref{fig:I}.

\begin{figure}[H]
\centering
\includegraphics[width=\linewidth]{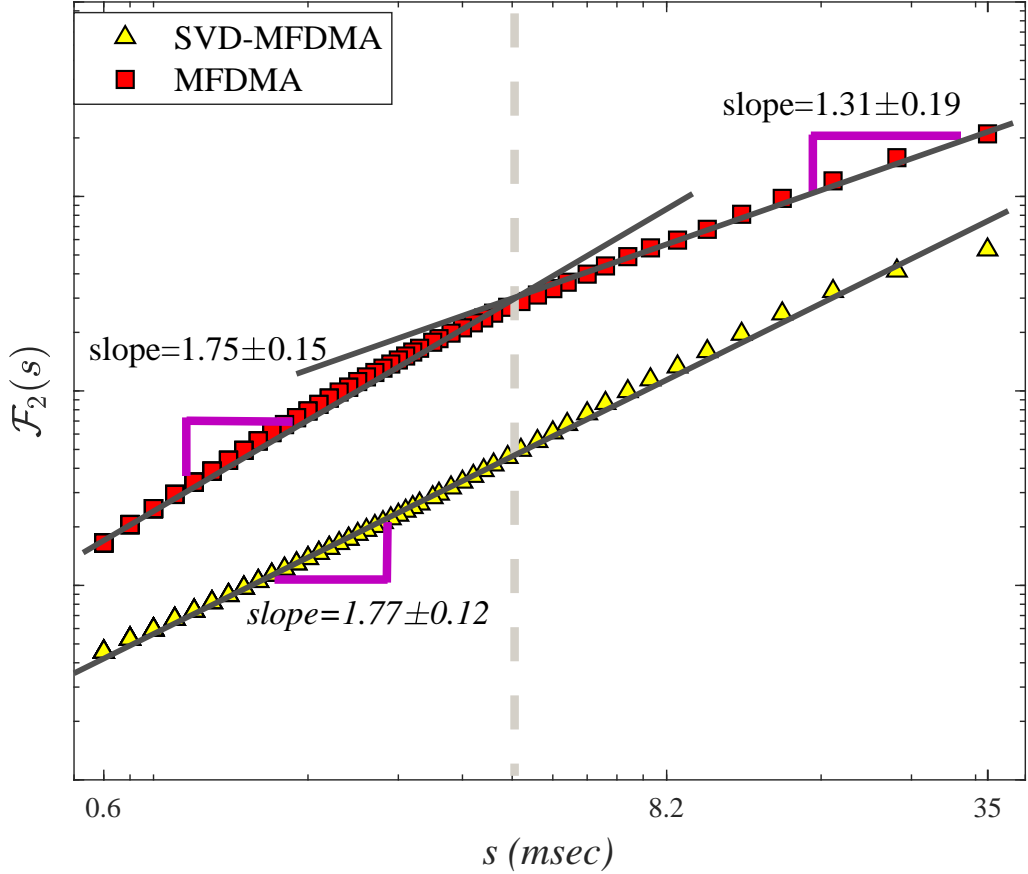}
\caption{{Log-Log plot of fluctuation function, $\mathcal{F}_2(s)$, as a function of $s$ for DMA, when we apply SVD as a pre-processor on a typical sABR data. The solid lines indicate the scaling behavior of fluctuation functions.}}
 \label{fig:dfacross}
 \end{figure}

 We use DMA on the observed sABR data sets. Our results demonstrate that there is a crossover time scale approximately equal to the 128 Hz in the fluctuation function versus scale. This crossover corresponds to $s_{\times}\sim7.8$ $ms$ and it is associated with the fundamental frequency of spoken syllable /da/. The scaling exponent for $s<s_{\times}$ is $h(q=2)=1.75\pm0.15$ demonstrating non-stationary nature of time series in this regime (Fig. \ref{fig:dfacross}). The slope of fluctuation function versus scale for $s>s_{\times}$ is $h(q=2)=1.31\pm0.19$. As illustrated in Fig. \ref{fig:dfacross}, SVD as the pre-processing algorithm can almost remove crossover which is almost devoted to the fundamental frequency. It is worth noting that mentioned behavior has been obtained for all $q$ values as well as for other sABR samples used in this paper.

\begin{figure}
\centering
\includegraphics[width=\linewidth]{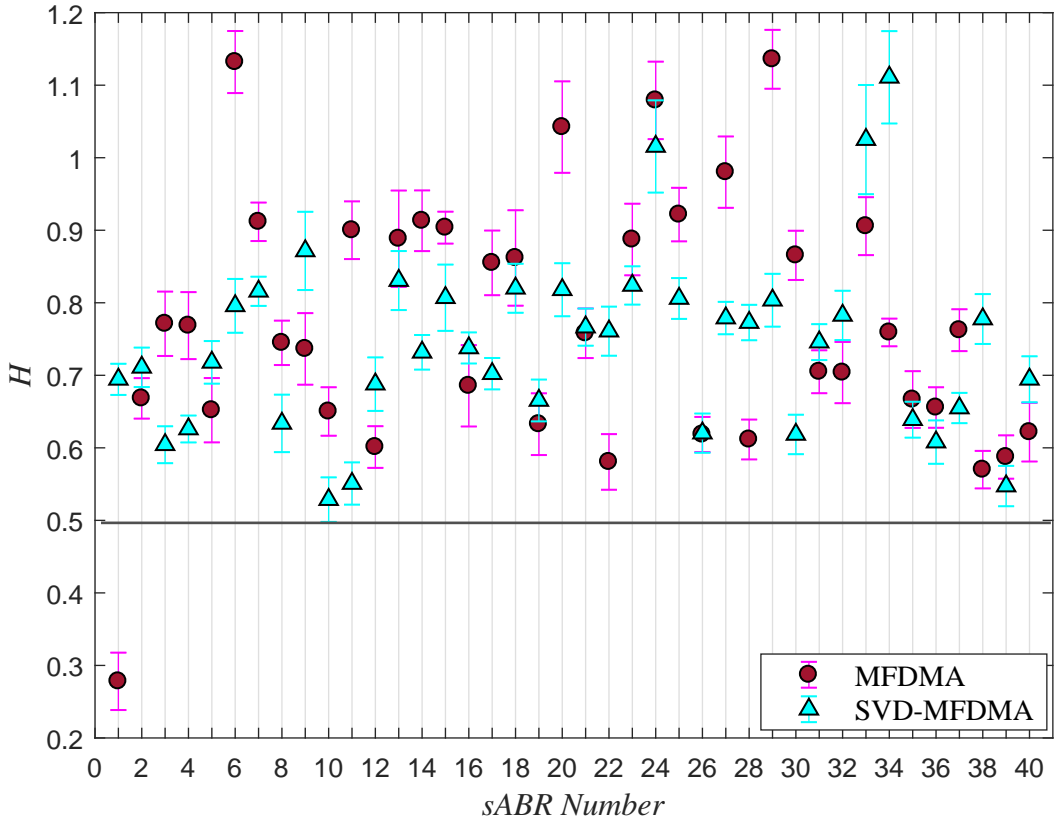}[H]
\caption{{The values of Hurst exponent computed by DMA for $s<s_{\times}$  and SVD-MFDMA methods for normal sABR series. The error-bar is reported at $1\sigma$ confidence level.}}
 \label{fig:hurstall}
 \end{figure}

In Fig. \ref{fig:hurstall}, we illustrate values of Hurst exponent determined by SVD-MFDMA for $q=2$ for all normal sABR series including associated error-bar at 68\% level of confidence. Our results show that Hurst exponent determined by SVD-MFDMA for $q=2$ for all normal sABR series is $H>0.5$ leading to long-range correlation behavior (equation (\ref{eq:h})). The slope of $\mathcal{F}_2$ for small $s$ is dominated by intrinsic fluctuations and consequently, it can be considered as a robust value for Hurst exponent \cite{hu2001effect,wei2017multifractal,li2018effect}.

\begin{figure}[H]
\centering
\includegraphics[width=\linewidth]{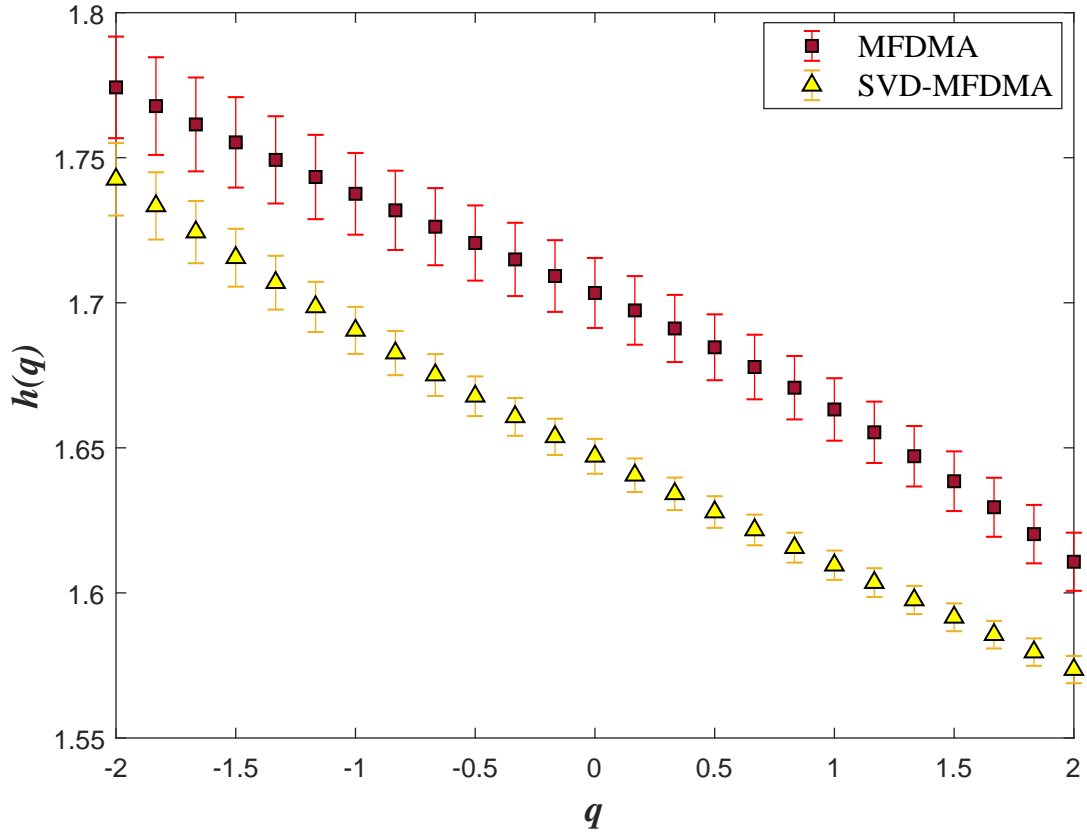}
\caption{{The $q-$dependency of the average $h(q)$ when we apply MFDMA and SVD-MFDMA methods for 40 normal sABR signals}}
\label{fig:fh}
\end{figure}

The generalized Hurst exponents averaged on 40 normal sABR signals applied by the MFDMA method for $s<s_{\times}$ is shown in Fig.  \ref{fig:fh} (square symbols). For $q>0$ the larger fluctuations have a dominant contribution in equation (\ref{eq:fluctuation function}), while small fluctuations are magnified for $q<0$. For a monofractal process, both mentioned behavior are similar leading to have constant $h(q)$. In another word, any $q-$dependency of generalized Hurst exponent reveals the multifractal nature of the underlying process. The presence of $q$ value enables us to manipulate the contribution of noises and trends in the underlying signal. Triangle symbols in Fig. \ref{fig:fh} correspond to generalized Hurst exponent averaged on 40 normal sABR samples computed by SVD-MFDMA. The values of generalized Hurst exponent by MFDMA for $s<s_{\times}$ are higher than that of computed by SVD-MFDMA. This result confirms that superimposed trends on data affect the value of fluctuations function even for $s<s_{\times}$, where we expect to obtain the nature of intrinsic fluctuations. Subsequently, to find reliable scaling exponents, we should implement a pre-processing method to reduce the effect of trends in sABR series.

\begin{figure}[H]
\centering
\includegraphics[width=\linewidth]{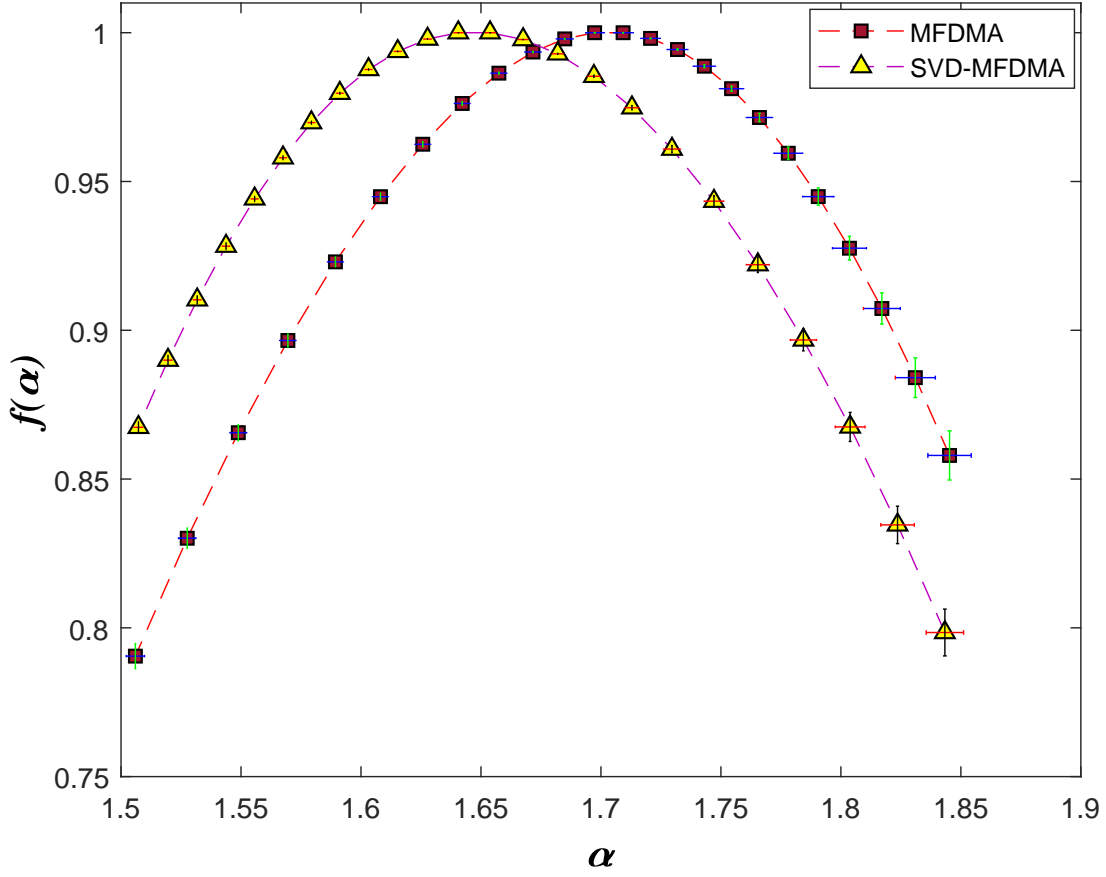}
\caption{{The singularity spectrum $f(\alpha)$ vs $\alpha$ when we apply MFDMA and MFDMA-SVD methods averaging on 40 normal sABR signals}}
\label{fig:falpha}
\end{figure}

The singularity spectrum $f(\alpha)$ of observed sABR series are shown in Fig. \ref{fig:falpha} (equation (\ref{eq:falpha})).

\begin{figure}[H]
\centering
\includegraphics[width=\linewidth]{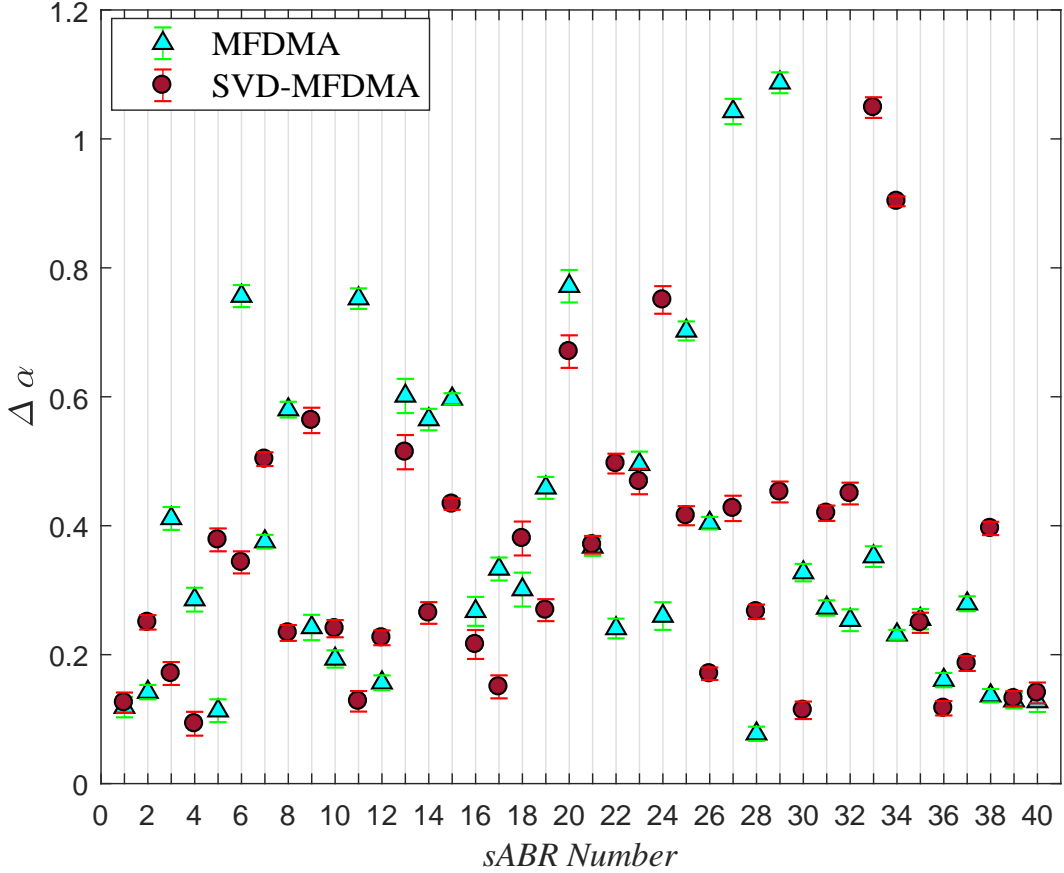}
\caption{{The values of the complexity measure, $\Delta\alpha$, computed by MFDMA and SVD-MFDMA methods for normal s-ABR series. The error-bar is reported at $1\sigma$ confidence level.}}
\label{fig:figcomp}
\end{figure}

The strength of multifractality nature of sABR is examined by width of singularity spectrum, $\Delta \alpha=\alpha_{\rm max}-\alpha_{\rm min}$. The values of  $\Delta \alpha$ for all normal sABR series including associated error-bar at 68\% level of confidence value are shown in Fig. \ref{fig:figcomp}.

\begin{figure}[H]
\centering
\includegraphics[width=\linewidth]{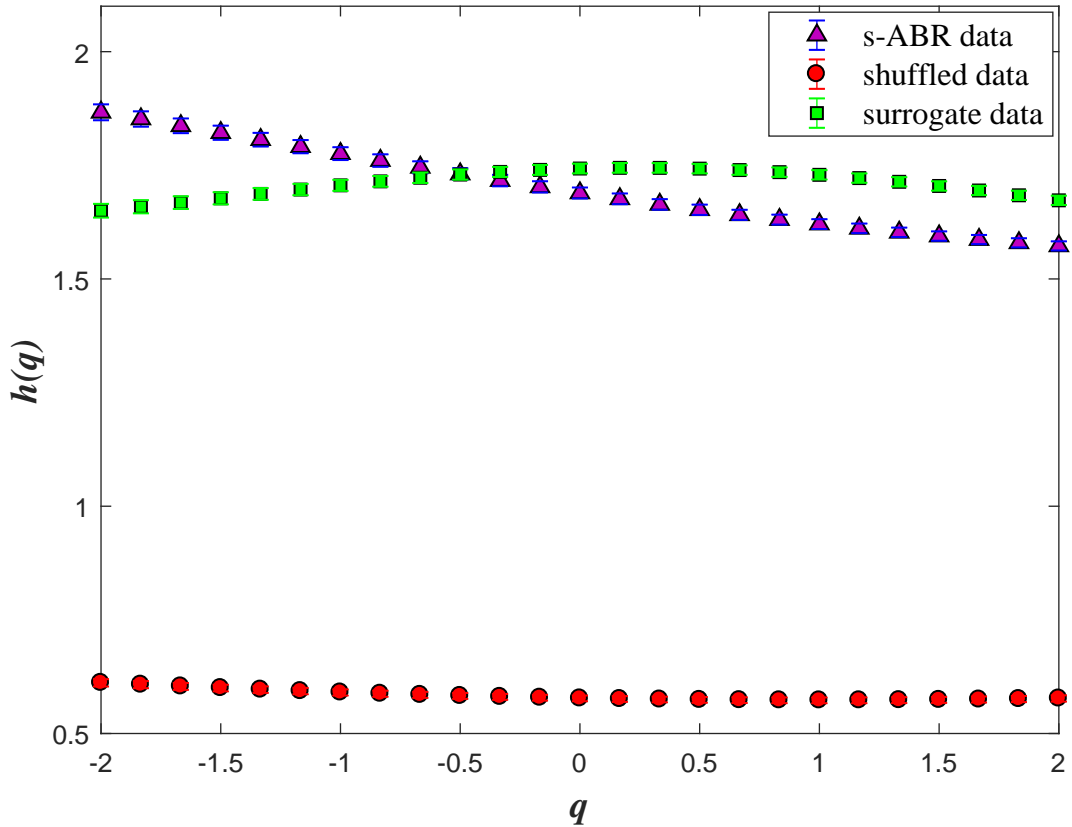}
\caption{{Generalized Hurst exponent, $h(q)$, of the original, surrogate and shuffled sABR signals averaging on 40 normal individuals.}}
\label{fig:SD}
\end{figure}

Now, we compare the fluctuation function of original sABR with the results of the corresponding shuffled and surrogate series to determine the nature of multifractality. In Fig. \ref{fig:SD}, the $q$ dependence of the $h(q)$ are shown for original, shuffled and surrogate series averaged on all normal sABR data sets. The sABR series values have been randomly shuffled to destroy the long-range correlations in the data. The shuffled data behaves as a monofractal signal and the $h(q)$ does not change in general with $q$ (Fig. \ref{fig:SD}). The sABR series have long-range correlations in different scales as it is obvious from the variance of $h(q)$ values corresponding to different $q$s. The presence of the long-range power-law correlations exhibit the fractal dynamics of the under investigation system \cite{alves2017long}.

The values of the Hurst exponent $h(q =2)$ for original, surrogate and shuffled sABR signal with MFDMA method averaging on 40 normal individuals are indicated in Table \ref{tab:sd}.

\begin{table}[H]
\centering
\begin{tabular}{|l|l|l|l|}
\hline
& \qquad sABR & \qquad  surrogate data & \qquad shuffled data \\
\hline
MFDMA & $1.75\pm0.15$ & $1.57\pm0.06$&$0.53\pm0.06$ \\
\hline
SVD-MFDMA & $1.77\pm0.18$ & $1.62\pm0.12$ &$0.50\pm0.09$\\
\hline
\end{tabular}
\caption{The values of the Hurst exponent $h(q =2)$ for original, surrogate and shuffled sABR signal with MFDMA method averaging on 40 normal people.}
\label{tab:sd}
\end{table}

\section{Discussion}\label{dis}
In this study, we have examined the large-scale dynamics of neural oscillations in the normal auditory system in the brainstem level. The presence of the long-range power-law correlations exhibit the fractal dynamics of the under-investigation system \cite{alves2017long}. Long-range temporal correlation has been obtained in a wide range of complex biological systems, including DNA sequences \cite{sutthibutpong2016long}, heart rate \cite{bhaduri2016quantitative}, medullary sympathetic neurons in neurophysiology \cite{lewis2001long, richard2018integrate}, human brain oscillations \cite{colombo2016more} and long memory in human coordination \cite{ducharme2018fractal}. All the studies lend considerable credence to the argument that an intrinsic part of the mechanism of neural information processing is the scale-free temporal correlation \cite{borges2017scale}. Various studies have revealed long-range temporal correlations in the neural oscillations of the normal human brain reflecting a memory of the underlying dynamics  \cite{dimitriadis2016modulation, alves2017long, meisel2017interplay}. In the present study, we have used the non-invasively recorded sABR to examine whether such scaling behavior occurs in the subcortical auditory structure.

We proposed a pipeline containing a combination of multifractal analysis (MFDMA) with singular value decomposition (SVD) to determine the multifractal characterization of the sABR series in the presence of trends. Our results based on the MFDMA, which does not have the discontinuity like the MFDFA method, reveal a cross-over, $s_{\times}\sim 7.8$ msec approximately equal to the fundamental frequency in the fluctuation functions versus scale for all sABR samples. The slope of $\mathcal{F}_2(s)$ as a function of $s$ averaged on all sABR series for $s<s_{\times}$ is $1.75 \pm 0.15$, while this slope for $s>s_{\times}$ is $1.31\pm0.19$ demonstrating that MFDMA cannot remove sinusoidal trends which are almost devoted to the fundamental frequency (Fig. \ref{fig:dfacross}). To determine the reliable generalized Hurst exponents of normal sABR signals, we applied SVD as a pre-processing algorithm to remove sinusoidal trends (Fig. \ref{fig:dfacross}). After applying SVD on our data sets, the MFDMA showed self-similarity features in normal sABR series. Also, the self-similarity parameters attained with the MFDMA method for $s<s_{\times}$ and by SVD-MFDMA for all available scales were invariant across subjects revealing a universality class for normal sABR samples subjected to the long-range temporal correlated signal. The presence of the $q-$dependency of $h(q)$ and $\Delta\alpha\neq0$ indicate the presence of robust multifractality in the sABR series corresponding to the second universality property. The width of the singularity spectrum on average equates to $\Delta\alpha=0.36\pm0.22$ at $1\sigma$ confidence interval quantifying the amount of multifractality of sABR series. The results demonstrate the complex structure of the sABR series revealing the presence of long-range correlation, which has been related to the slow timescales in subcortical level and integration of information in brainstem network \cite{favela2016fractal}.

We have also found the source of multifractality in the sABR is almost related to the long-range temporal correlations confirmed by comparing generalized Hurst exponent of original sABR series with the results of the corresponding shuffled which is devoted to the third universality. This means that we have a generic property in all data sets from the source of multifractality point of view. By comparing the scaling exponents of original data with that of computing for surrogate series, we obtained that the contribution of probability distribution function broadness in complexity behavior is not significant. Our findings exhibited that fluctuations in the sABR series are dictated by a mechanism associated with long-term memory of the dynamic of the auditory system in the brainstem level revealing third universality class \cite{alves2017long}. This evaluation provides deeper insight into the time series originating from the auditory system in the brainstem level. Further studies are required to examine how the long-range correlation is affected by different pathological conditions and evaluate the hypothesis that the MFDMA analysis of sABR series would reveal the difference in the long-range temporal correlation between pathological and normal conditions quantitatively. This pipeline can be used as a pre-processor to remove different types of noises and trends and also evaluation of other features of underlying sABR series. Evaluation of this approach as a diagnostic method is in progress.

\vspace{1 cm}

\textbf{Acknowledgments:}
The authors thank Mohsen Ahadi for collecting the data. This work was partially supported by the Iran National Science Foundation (INSF) through the Grant number 96010213.

\end{document}